\documentclass[9pt,conference]{IEEEtran}
\usepackage{fancyhdr}
\usepackage{amsmath} 
\usepackage{mathtools} 
\usepackage{algorithm}
\usepackage{algorithmic}
\usepackage{stmaryrd}
\usepackage{amsfonts}

\usepackage[preprint]{waspaa25}

\usepackage{bm} 
\usepackage{nicefrac}
\usepackage{cleveref} 

\def\vs{{{s}}}
\def\vphi{{{\phi}}}
\def\rvx{{{x}}}

\newcommand{\mcal}[1]{\mathcal{#1}}
\def\mI{{{I}}}
\def\vtheta{{{\theta}}}
\def\vf{{{f}}}

\title{Robust One-step Speech Enhancement via Consistency Distillation}

\name{Liang Xu$^{1}$,
      Longfei Felix Yan$^{1}$,
      W. Bastiaan Kleijn$^{1}$\thanks{Accepted to the 2025 IEEE WASPAA}}
\address{$^{1}$School of Engineering and Computer Science, Victoria University of Wellington, Wellington 6140, New Zealand \\
\{liang.xu, felix.yan, bastiaan.kleijn\}@vuw.ac.nz
}
\begin{document}

\maketitle

\begin{abstract}
Diffusion models have shown strong performance in speech enhancement, but their real-time applicability has been limited by multi-step iterative sampling. Consistency distillation has recently emerged as a promising alternative by distilling a one-step consistency model from a multi-step diffusion-based teacher model. However, distilled consistency models are inherently biased towards the sampling trajectory of the teacher model, making them less robust to noise and prone to inheriting inaccuracies from the teacher model. To address this limitation, we propose ROSE-CD: Robust One-step Speech Enhancement via Consistency Distillation, a novel approach for distilling a one-step consistency model. Specifically, we introduce a randomized learning trajectory to improve the model's robustness to noise. Furthermore, we jointly optimize the one-step model with two time-domain auxiliary losses, enabling it to recover from teacher-induced errors and surpass the teacher model in overall performance. This is the first pure one-step consistency distillation model for diffusion-based speech enhancement, achieving 54 times faster inference speed and superior performance compared to its 30-step teacher model. Experiments on the VoiceBank-DEMAND dataset demonstrate that the proposed model achieves state-of-the-art performance in terms of speech quality. Moreover, its generalization ability is validated on both an out-of-domain dataset and real-world noisy recordings.

\end{abstract}

\section{Introduction}
\label{sec:intro}

Speech enhancement (SE), the task of recovering clean speech from noise-contaminated signals, is fundamental to robust speech communication. Classical SE approaches include Wiener filtering, e.g.,~\cite{meyer1997multi}, and beamforming~\cite{benesty2017fundamentals, chua2024effective,yan2025neural}. While effective in certain conditions, these methods often degrade in highly non-stationary environments or rely on the spatial settings of microphone arrays. Recent advances in data-driven SE have led to the development of predictive, generative, and hybrid models. Predictive models typically learn a deterministic mapping from noisy to clean speech, producing an estimate of the clean signal~\cite{su2021hifi, strauss2023sefgan, fu21_interspeech, luo2019conv}. In contrast, generative models aim to learn the conditional distribution of clean speech given noisy input, enabling more diverse and robust outputs across varying noise conditions~\cite{lu2022conditional, richter2025investigating, richter2023speech, jukic24_interspeech, CRP2024Lay}. Hybrid approaches~\cite{lemercier2023storm, trachu24_interspeech, shi2024diffusion} attempt to combine the strengths of predictive and generative models to further enhance robustness, often at the cost of additional computational overhead.

Recently, diffusion-based generative models have demonstrated state-of-the-art performance in speech enhancement~\cite{lu2022conditional, richter2023speech, jukic24_interspeech, richter2025investigating, shi2024diffusion}. Nevertheless, their reliance on iterative multi-step reverse diffusion processes remains a major obstacle for real-time deployment. For example, methods such as CDiffuSE~\cite{lu2022conditional} and SGMSE+~\cite{richter2023speech} typically require 30 to 200 inference steps to reconstruct clean speech, resulting in substantial computational overhead and latency.

To address the computational limitation, recent research has focused on reducing the number of denoising steps to enhance inference efficiency. For example, CRP~\cite{CRP2024Lay} proposes a two-stage training scheme, introduces a predictive loss in the second stage to fine-tune the score model, achieving good performance with only 5 reverse steps. The hybrid approach StoRM~\cite{lemercier2023storm} introduces a predictive model before diffusion, enabling sampling schemes with fewer diffusion steps without sacrificing quality. Furthermore, Thunder~\cite{trachu24_interspeech} extends StoRM by incorporating a Brownian bridge process~\cite{karatzas1991brownian}, enabling a flexible fusion strategy that integrates a regression model with a one-step diffusion model. However, this fusion mechanism adds computational overhead, resulting in a two-step inference process.

Another promising approach for reducing sampling steps in diffusion models is consistency models~\cite{pmlr-v202-song23a}, which enable efficient one-step generation without adversarial training. Specifically, consistency models can be trained via two distinct strategies: direct consistency training (CT)~\cite{qiu2023sebridge} or consistency distillation (CD) from a pre-trained diffusion teacher model. It has been demonstrated that CD often delivers better performance than direct CT~\cite{pmlr-v202-song23a}. This is primarily because CD leverages a pre-trained diffusion teacher model to provide a high-quality score function, which helps reduce variance in the training loss and enables more stable optimization. Consequently, this can lead to superior sample quality and faster convergence compared to CT, whose standalone training relies on a potentially noisier score estimator, introducing higher variance and bias. Given these advantages, consistency distillation has become a preferred method for achieving efficient and high-fidelity speech enhancement.

We present ROSE-CD: Robust One-step Speech Enhancement via Consistency Distillation, a framework that achieves the fastest inference speed while surpassing its 30-step teacher in speech enhancement quality. ROSE-CD improves model robustness and overall performance through two key innovations. First, it introduces randomized trajectory learning during distillation, which enhances robustness and mitigates overfitting to teacher-induced biases. Second, it incorporates two time-domain auxiliary losses—PESQ and SI-SDR—to facilitate direct learning from clean data distributions and improve recovery from teacher model errors. In our framework, the teacher model serves primarily as a reference, while the one-step consistency model learns a robust and accurate representation by leveraging both randomized trajectories and time-domain clean signals. As a result, we achieve a 54$\times$ speed-up in inference over its teacher model and attain a state-of-the-art PESQ score of 3.99 on the VoiceBank-DEMAND dataset. Furthermore, extensive evaluations confirm its strong generalization capability on both an out-of-domain dataset and real-world noisy recordings.

\section{METHODOLOGY}
\label{sec:METHODOLOGY}

Inspired by consistency models~\cite{pmlr-v202-song23a} and recent advances in joint learning strategies~\cite{deoliveira2024pesqetarian}, we propose to distill a robust one-step consistency model from a 30-step diffusion-based teacher model. This section begins with a review of score-based diffusion models, followed by a detailed description of the proposed robust consistency distillation method and the joint optimization strategy.

\subsection{Score-based Diffusion Model for Speech Enhancement}
\subsubsection{Preliminaries}
\label{ssec:Preliminaries}

Score-based diffusion models~\cite{song2019generative, song2021scorebased} define two processes: the forward process and the reverse process. The forward process involves the gradual addition of noise and is described by the solution to the following stochastic differential equation (SDE):

\begin{align} \label{eq:forward_sde}
    \mathrm{d} x_t = f(x_t, y) \mathrm{d} t + g(t) \mathrm{d} w,
\end{align}
where $f(x_t, y)$ and $g(t)$ represent the drift and diffusion coefficients, respectively. The variables $x_t$, $y$, and $w$ denote the state of the process at time $t \in [0, 1]$, the noisy speech signal, and a standard Wiener process, respectively. Similarly, the reverse SDE is given by:
\begin{align} \label{eq:reverse_sde}
    \mathrm{d} x_t = \left[ f(x_t, y) - g(t)^2 \nabla_{x_t} \log p_t(x_t | y) \right] \mathrm{d} t + g(t) \mathrm{d} \bar{w},
\end{align}
where $\nabla_{x_t} \log p_t(x_t | y)$ is the conditional score function, and $\bar{w}$ is a standard Wiener process evolving backward in time.

In ~\cite{richter2023speech} and ~\cite{richter2025investigating}, they use a drift coefficient of the form $f(x_t, y) = \gamma (y - x_t)$ where the stiffness parameter $\gamma$ controls the rate of transformation from $x_0$ to $y$. They furthermore select a diffusion coefficient $g(t) = \sqrt{c} k^{t}$ with positive parameters $c$ and $k$. The conditional transition distribution is described by the perturbation kernel:
\begin{align}
\label{eq:perturbation-kernel}
    p_{t}(x_t|x_0, y) = \mathcal{N}_\mathbb{C}\left(x_t; \mu(x_0, y, t), \sigma(t)^2 \mathbf{I}\right),
\end{align}
where $\mathcal{N}_\mathbb{C}$ denotes the circularly symmetric complex normal distribution. The mean $\mu(x_0, y, t)$ and variance $\sigma(t)^2$ are given by:
\begin{align}
\label{eq:mean}
    \mu(x_0, y, t) = \mathrm{e}^{-\gamma t} x_0 + (1 - \mathrm{e}^{-\gamma t}) y, \enspace\thickspace
    \sigma(t)^2 = \frac{c \left( k^{2t} - \mathrm{e}^{-2 \gamma t} \right)}{2 \left( \gamma + \log k \right)}.
\end{align}
\subsubsection{Optimizing Goals} 
\label{ssec:Optimizing}

Direct computation of $\nabla_{x_t} \log p_t(x_t \mid y)$ is generally intractable. To circumvent this, following the approaches in~\cite{richter2023speech, song2021scorebased}, we train a score model $s_\theta(x_t, y, t)$ to approximate the conditional score function $\nabla_{x_t} \log p_t(x_t \mid x_0, y)$, employing a denoising score matching objective:
\begin{align} \label{eq:dsm_loss}
\mathcal L_\text{score} = \lambda(t)\Big|\Big|s_\theta(x_t, y, t) + \frac{z}{\sigma(t)}\Big|\Big|_2^2,
\end{align}
where $t$ is randomly sampled from $\mathcal{U}[0, 1]$, and $\lambda(t)$ is a weighting function, $x_t$ is sampled from the perturbed distribution $p_t(x_t|x_0, y)$, $z \sim \mathcal{N} (0, I)$ is the random noise.

As shown in~\cite{hyvarinen2005estimation, vincent2011connection} and widely adopted in modern diffusion models~\cite{karras2022elucidating}, minimizing the score matching objective is equivalent to training a denoiser model $D_\theta(x_t, y, t) = x_t + \sigma_t^2 \cdot s_\theta(x_t, y, t)\,$with the following denoising loss:
\begin{equation}
\label{eq:denoising_loss}
 \mathcal{L}_\text{denoise} = \lambda(t) \lVert D_\theta(x_t, y, t) - \mu_t(x_0, y) \rVert^2_2.
\end{equation}
Empirically, it is beneficial to parameterize the denoiser $D_\theta$ using skip connections:
\begin{equation}
\label{eq:preconditioning}
    D_\theta (x_t, y, t) = c_\text{skip}(t) x_t + c_\text{out}(t) F_\theta (c_\text{in}(t) x_t, c_\text{in}(t) y, t),
\end{equation}
where $c_\text{skip}(t)$, $c_\text{out}(t)$, and $c_\text{in}(t)$ are time-dependent scaling functions derived in~\cite{karras2022elucidating}, satisfying the boundary conditions $c_\text{skip}(0) = 1$ and $c_\text{out}(0) = 0$. Here, $F_\vtheta(\rvx, y, t)$ denotes a neural network that produces outputs with the same dimensionality as input $\rvx$, which normally shares the same parameters with $s_\theta$.

After estimating the conditional score function $\nabla_{x_t} \log p_t(x_t | y)$ for all time steps $t$, the corresponding reverse-time SDE in~\eqref{eq:reverse_sde} can be used to generate clean speech by denoising $y$. It has been shown in~\cite{richter2025investigating} that training a diffusion-based speech enhancement model using either $\mathcal{L}_\text{denoise}$ or $\mathcal{L}_\text{score}$ leads to comparable performance.

\subsection{Robust Consistency Distillation}
\label{ssec:RCD}

\subsubsection{Consistency Distillation}
\label{ssec:CD}

Consistency distillation~\cite{pmlr-v202-song23a} aims to distill a one-step consistency model $f_\theta(x_t, y, t)$ from a pre-trained multi-step teacher model $\vs_\vphi(\rvx,y, t)$, which is a diffusion-based score model that defines the ODE trajectory used in the backward process. Specifically, we divide the discrete time interval $[\delta, T]$ into $N - 1$ sub-intervals and randomly sample a state $\rvx_{t_n}$ according to the perturbation kernel defined in~\eqref{eq:perturbation-kernel}, where $\delta$ is a small positive constant (e.g., $\delta=0.03$) introduced to avoid numerical instability. The goal is to estimate the preceding state $\rvx_{t_{n-1}}$ using a one-step ODE solver:
\begin{align}
    \hat{\rvx}_{t_{n-1}}^\vphi = \rvx_{t_{n}} + (t_{n-1} - t_{n})\Phi(\rvx_{t_{n}},y, t_{n}; \vphi),
    \label{eq:odesolve}
\end{align}
where $\Phi(\cdot; \vphi)$ denotes the update function of a one-step ODE solver. We define the consistency distillation loss as: 
\begin{equation}
    \mathcal{L}_\text{CD}^N(\vtheta, \vtheta^{-}; \vphi) = \mathbb{E}[\lambda(t_{n-1}) d(\vf_\vtheta(\rvx_{t_{n}}, y, t_{n}), \vf_{\vtheta^-}(\hat{\rvx}_{t_{n-1}}^\vphi, y, t_{n-1}))],
    \label{eq:distill_obj}
\end{equation}
where $d(\cdot)$ denotes the distance function, $\lambda(\cdot) \in \mathbb{R}^+$ is a weighting function, and $\vtheta^{-}$ represents the exponential moving average (EMA) of the historical parameters of $\vtheta$. Following the setup in~\cite{pmlr-v202-song23a}, we employ the $L_2$ distance as $d(\cdot)$ and set $\lambda(\cdot) = 1$. The consistency model $f_\theta(x_t, y, t)$ is parameterized using a skip-connection architecture:
\begin{align}
    \vf_\vtheta(x_t, y, t) = d_\text{skip}(t) x_t + d_\text{out}(t) F_\vtheta(x_t, y, t), \label{eq:param2}
\end{align}
where $d_\text{skip}(t)$ and $d_\text{out}(t)$~\cite{karras2022elucidating} are differentiable weighting functions satisfying $d_\text{skip}(0) = 1$ and $d_\text{out}(0) = 0$, and $f_\theta$ eliminates the need for input scaling~\cite{pmlr-v202-song23a}.

\subsubsection{Robust Consistency Distillation}
\label{ssec:RRCD}
\begin{figure}[!t]
  \centering
  \centerline{\includegraphics[width=\columnwidth]{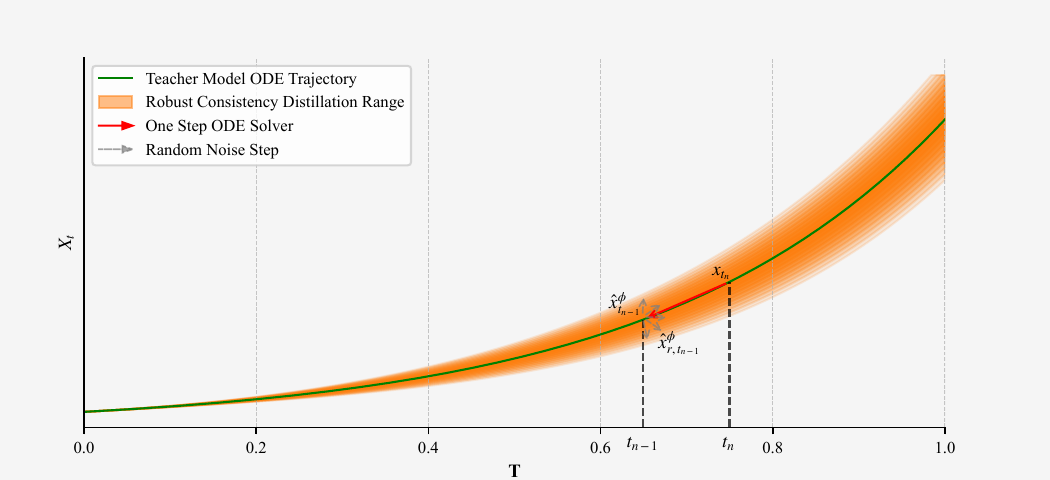}}
  
  \caption{Overview of the proposed robust consistency distillation (RCD). The thick green line illustrates the PF-ODE trajectory defined by a pre-trained diffusion teacher model. During distillation, given a sampled data point $\rvx_{t_n}$ at time step $t_n$, we first estimate $\hat{\rvx}_{t_{n-1}}^\vphi$ using a one-step ODE solver. To improve robustness, a random noise perturbation is then applied to obtain a noised variant $\hat{\rvx}_{r,t_{n-1}}^\vphi$. Finally, the consistency model is trained within this robust consistency distillation range, which is highlighted in orange.
  }
  \label{figure:results}
\end{figure}

As illustrated in~\Cref{figure:results}, given a multi-step diffusion-based teacher model $\vs_\vphi(\rvx,y, t)$, we train the consistency model $f_\theta(\rvx_t, y, t)$ on adjacent time-step pairs such that it satisfies the following condition:
\begin{equation}
    \vf_\vtheta(\rvx_{t_{n}}, y, t_{n})=\vf_{\vtheta}(\hat{\rvx}_{t_{n-1}}^\vphi, y, t_{n-1}),
    \label{eq:cd_equation}
\end{equation}
where the estimated $\hat{\rvx}_{t_{n-1}}^\vphi$ is obtained using the teacher model’s ODE trajectory, as described in~\eqref{eq:odesolve}.

We inject an additional noise term into the ODE-based trajectory estimation to prevent the consistency model from merely imitating potentially flawed teacher trajectories. In standard consistency distillation, the teacher’s trajectory is fully deterministic for a given \(\rvx_{t_n}\), leading the consistency model to eventually imitate the teacher’s behavior once training converges. However, this strict alignment can limit the robustness and generalization ability of the consistency model, as it inevitably inherits the teacher’s errors and biases. To address this, we modify the one-step estimation process as follows:
\begin{align}
    \hat{\rvx}_{r,t_{n-1}}^\vphi = \rvx_{t_n} + (t_{n-1} - t_n) \Phi(\rvx_{t_n}, y, t_n; \vphi) + g(t) \sqrt{\Delta t} \, \boldsymbol{\epsilon},
    \label{eq:robustodesolve}
\end{align}
where \(g(t)\) is the diffusion coefficient defined in~\eqref{eq:forward_sde}, \(\boldsymbol{\epsilon} \sim \mathcal{N}(0, I)\) denotes the added noise, and \(\Delta t = t_n - t_{n-1}\). This perturbation compels the consistency model to learn from noisy adjacent pairs \((\hat{\rvx}_{r,t_{n-1}}^\vphi, \rvx_{t_n})\), thereby enhancing its robustness to noise.

\begin{algorithm}[!htbp]
    \caption{Robust Consistency Distillation (RCD)}\label{alg:distillation}
 \begin{algorithmic}
    \STATE {\bfseries Input:} dataset $\mcal{D}$, initial consistency model parameter $\vtheta$, learning rate $\eta$, ODE solver $\Phi(\cdot, \cdot; \vphi)$, $d(\cdot, \cdot)$, EMA decay rate $\mu$, and diffusion coefficient $g(\cdot)$
    \STATE $\vtheta^- \gets \vtheta$
    \REPEAT
    \STATE Sample $x_0,y \sim \mcal{D}$ and $n \sim \mcal{U}\llbracket 2, N \rrbracket$
    \STATE Sample $\rvx_{t_{n}} \sim p_{t}(x_t|x_0, y)$ by~\eqref{eq:perturbation-kernel}
    \STATE Sample $\boldsymbol{\epsilon} \sim \mcal{N}(0, \mI)$
    \STATE  $\hat{\rvx}_{r,t_{n-1}}^\vphi = \rvx_{t_{n}} + (t_{n-1} - t_{n})\Phi(\rvx_{t_{n}},y, t_{n}; \vphi) + g(t)\sqrt{\Delta t
} \, \mathbf{\boldsymbol{\epsilon}}$
    \vspace{0.33em}
    \STATE $\begin{multlined}
        \mcal{L}_\text{RCD}(\vtheta, \vtheta^{-}; \vphi) \gets d(\vf_\vtheta(\rvx_{t_{n}}, y, t_{n}), \vf_{\vtheta^-}(\hat{\rvx}_{r,t_{n-1}}^\vphi, y, t_{n-1}))
    \end{multlined}
    $
    \vspace{0.33em}
    \STATE $\vtheta \gets \vtheta - \eta \nabla_\vtheta \mcal{L}_\text{RCD}(\vtheta, \vtheta^{-}; \vphi)$
    \STATE $\vtheta^- \gets \operatorname{stopgrad}(\mu \vtheta^- + (1-\mu) \vtheta$)
    \UNTIL{convergence}    
 \end{algorithmic}
 \end{algorithm}
 
~\Cref{alg:distillation} describes the training process of RCD. After the robust consistency model $f_\theta$ is well trained, clean speech can be generated directly through a single reverse step $x=f_\theta(\rvx_{T},y, T)$, starting from the initial noisy sample $\rvx_{T} \sim \mathcal{N}_\mathbb{C}\left(y, \sigma(T)^2 \mathbf{I}\right)$.

\subsubsection{Joint Optimization}
\label{ssec:joint}

Drawing inspiration from~\cite{richter2025investigating}, we propose a joint optimization strategy for the one-step consistency model, incorporating two auxiliary time-domain losses:
\begin{equation}
    \mathcal{L} = \mathcal{L}_{\text{RCD}} + \lambda_1 \mathcal{L}_{\text{PESQ}} \left( \underline{\hat{\mathbf{x}}}_\theta(t_n), \underline{\mathbf{x}}_0 \right) + \lambda_2 \mathcal{L}_{\text{SI-SDR}} \left( \underline{\hat{\mathbf{x}}}_\theta(t_n), \underline{\mathbf{x}}_0 \right),
    \label{eq:joint_loss}
\end{equation}
where $\mathcal{L}_{\text{RCD}}$ represents the robust consistency distillation loss in~\Cref{alg:distillation}. For $\mathcal{L}_{\text{PESQ}}$, we adopt the differentiable implementation from torch-pesq~\footnote{\url{https://github.com/audiolabs/torch-pesq}}, which builds upon~\cite{PESQ_loss_adeep, kim2019end}. For $\mathcal{L}_{\text{SI-SDR}}$, we employ the negative SI-SDR loss as defined in~\cite{le2019sdr}. The hyperparameters $\lambda_1$ and $\lambda_2$ control the weighting of the PESQ and SI-SDR losses, respectively. The time-domain signals $\underline{\hat{\mathbf{x}}}_\theta(t_n)$ and $\underline{\mathbf{x}}_0$ are obtained via the inverse short-time Fourier transform (iSTFT), where $\underline{\hat{\mathbf{x}}}_\theta(t_n) = \mathrm{iSTFT}(\vf_\vtheta(\rvx_{t_n}, y, t_n))$ is the predicted waveform and $\underline{\mathbf{x}}_0 = \mathrm{iSTFT}(x_0)$ is the ground-truth reference. This encourages the model to learn directly from clean data distributions, facilitates recovery from teacher-induced errors, and enhances both perceptual quality and temporal fidelity in the generated speech.

\section{EXPERIMENTAL SETUP}
\label{sec:EXPERIMENTAL}

\subsection{Dataset}
\label{ssec:Dataset}

Following previous studies~\cite{lemercier2023storm, richter2023speech}, we adopted the VoiceBank-DEMAND (VB-DMD) dataset~\cite{botinhao2016investigating, thiemann2013diverse}, which comprised recordings from 30 speakers in the VoiceBank corpus~\cite{botinhao2016investigating}, with 26 used for training and 2 each for validation and testing. The training and validation sets included 11,572 utterances corrupted with eight real-world noises from DEMAND~\cite{thiemann2013diverse} and two synthetic noises (babble, speech-shaped) at 0, 5, 10, and 15~dB SNR. The test set comprised 824 utterances mixed with different noise samples at 2.5, 7.5, 12.5, and 17.5~dB SNR. For fair comparison, we also resampled all audio samples with a sampling rate of 16 kHz.

To evaluate the model's generalization capability, we performed assessments on both an out-of-domain dataset and real-world recordings. For the former, we utilized the TIMIT+NOISE92 dataset, which was constructed by corrupting the 1344 utterances from the TIMIT complete test set~\cite{garofolo1993timit} with 15 real-world noise samples from the NOISE92 dataset~\cite{varga1993assessment}. These utterances were corrupted at SNR levels of 0, 5, 10, and 15~dB. For the real-world recordings evaluation, we utilized 300 test recordings from the Deep Noise Suppression (DNS) Challenge 2020~\cite{reddy2020interspeech}. These recordings consisted of real-world data collected internally at Microsoft, covering a variety of noisy acoustic conditions and captured using different devices, including headphones and speakerphones.

\subsection{Implementation Details}
\label{ssec:Implementation}
We conducted consistency distillation from the SGMSE+ model~\cite{richter2023speech}, which adapted a NCSN++V2 network~\cite{richter2025investigating} as a backbone in the spectral domain. To serve as the multi-step teacher model, we retrained a variant of SGMSE+ that was parameterized using skip connections, as proposed in EDM~\cite{karras2022elucidating} and detailed in~\eqref{eq:preconditioning}. Specifically, we adopted a default reverse time step of \(N=30\) for the distillation process, the loss weights for PESQ and SI-SDR were $\lambda_{1}=5\times10^{-4}$ and $\lambda_{2}=5\times10^{-5}$, respectively. The audio data preprocessing followed the original SGMSE+ configuration.

All models were trained on the VB-DMD dataset for up to 100 epochs using a single NVIDIA A40 GPU (48 GB RAM). We employed the Adam optimizer with a learning rate of $\eta=10^{-4}$, a batch size of 32, and an EMA decay rate of $\mu=0.9999$. The model checkpoint that achieved the highest PESQ score on the validation set was selected. During distillation, the teacher model's weights were used to initialize the consistency model and remained fixed throughout the entire distillation process.

\subsection{Evaluation Metrics}
\label{ssec:Evaluation}
We evaluated performance using both reference-based and reference-free metrics. The former compares enhanced speech to clean ground truth and was applied to both in-domain and out-of-domain scenarios. The latter uses deep neural networks for non-intrusive assessment without requiring clean references.
\subsubsection{Reference-based metrics}
\label{sssec:Reference_based_metrics}
We used PESQ~\cite{rix2001perceptual} for speech quality (1–4.5), ESTOI~\cite{jensen2016algorithm} for intelligibility (0–1), and SI-SDR~\cite{le2019sdr} to assess signal fidelity in dB, with higher values indicating better performance.

\subsubsection{Reference-free metrics}
\label{sssec:Reference_free_metrics}
We used WV-MOS~\cite{andreev2023hifi++} to estimate the Mean Opinion Score (MOS) for speech quality using a wav2vec2.0-based model~\cite{baevski2020wav2vec}, and DNSMOS~\cite{reddy2021dnsmos} to assess perceptual quality. DNSMOS P.835~\cite{reddy2022dnsmos} further provides three component scores: Speech Quality (SIG), Background Noise Quality (BAK), and Overall Quality (OVRL). For complete comparison, we also report MOS-SSL~\cite{cooper2022generalization}.

\section{Results}
\label{sec:results}

\subsection{In-domain evaluation}
\label{subsec:in_domain}

In~\Cref{tab:performance}, we compared our proposed model against several methods categorized into three groups: predictive models, pure generative models, and hybrid models. Our model consistently surpassed the 30-step teacher model across all metrics. Furthermore, unlike hybrid approaches that operated in a two-step manner by fusing a predictive component with a generative model, our method directly utilized a single reverse step. Notably, our model achieves the highest PESQ score, surpassing PESQetarian~\cite{deoliveira2024pesqetarian} as well as the M6 and M7 models from~\cite{richter2025investigating}. While CRP~\cite{CRP2024Lay} supports one-step generation, it suffers from degraded performance.

To evaluate the effectiveness of RCD, we conducted experiments using two ODE solvers: Euler (e.g.,~\cite{song2021scorebased}) and Heun (e.g.,~\cite{karras2022elucidating}). RCD consistently enhanced performance for both solvers and notably helped narrow the gap between them. With the Euler solver, the PESQ score improved from 2.46 to 2.88, and the SI-SDR increased from 14.30~dB to 18.30~dB, surpassing the baseline Heun solver without RCD. It is clear that using the Heun solver with RCD yielded the best PESQ and SI-SDR performance and also surpassed the teacher model in both metrics. Therefore, we adopted Heun as the default solver for all subsequent experiments.

To investigate the roles of auxiliary time-domain losses, we evaluated the impact of optimizing with PESQ and SI-SDR losses both individually and jointly. When optimizing solely with the PESQ loss, we achieved the highest PESQ score of 3.99, indicating a significant improvement in perceptual quality. However, this improvement came at the cost of a substantial degradation in SI-SDR to 0.40~dB, highlighting poor temporal alignment. This contrast underscores the distinct nature of the two metrics: PESQ focuses on perceptual quality and can tolerate time shifts, while SI-SDR demands strict temporal synchronization and penalizes even slight temporal variations, regardless of perceptual improvements. On the other hand, optimizing exclusively with the SI-SDR loss preserved strong performance in temporal alignment, with an SI-SDR score of 17.30, but resulted in only a modest improvement in PESQ. 

Our proposed ROSE-CD, which uses joint optimization with both PESQ and SI-SDR losses, effectively maintains both high perceptual quality and strong temporal fidelity, achieving a PESQ of 3.49, an SI-SDR of 17.80, a SOTA MOS-SSL of 4.13, and the second-best WV-MOS score of 4.41.

\begin{table}[!t]
\centering
\caption{Performance comparison on the VB-DMD test set. The best results within each section are highlighted in \textbf{bold}. Other existing methods are grouped by algorithm type: predictive (P), pure generative (G), or hybrid (P+G). For hybrid methods, the number of steps used in both the predictive and generative modules is specified. RCD refers to robust consistency distillation.}

\label{tab:performance}
\sisetup{
    reset-text-series = false, 
    text-series-to-math = true, 
    mode=text,
    tight-spacing=true,
    round-mode=places,
    round-precision=2,
    table-format=2.2,
    table-number-alignment=center
}
\resizebox{\linewidth}{!}{
\begin{tabular}{l c c S[table-format=1.2] S[table-format=1.2] S[table-format=2.2] S[table-format=1.2] S[table-format=1.1]}
    \toprule
    \textbf{Method} & \textbf{Type} & \textbf{Steps} & \textbf{PESQ} & \textbf{ESTOI} & \textbf{SI-SDR} & \textbf{MOS-SSL} & \textbf{WV-MOS} \\
    \midrule    
    MetricGAN+~\cite{fu21_interspeech} & P & -- & 3.13 & 0.83 & 8.50 & 3.25 & 3.56 \\
    PESQetarian~\cite{deoliveira2024pesqetarian} & P & -- & \textbf{3.82} & 0.84 & -19.80 & {--} & {--} \\
    StoRM~\cite{lemercier2023storm} & P+G & 51 & 2.93 & \textbf{0.88} & 18.80 & 4.09 & \textbf{4.26} \\
    Thunder~\cite{trachu24_interspeech} & P+G & 2 & 3.02 & 0.87 & \textbf{19.40} & \textbf{4.10} & 4.23 \\     
    \midrule
    M6~\cite{richter2025investigating} & G & 30 & \textbf{3.70} & 0.86 & 8.30 & 3.90 & 4.33 \\
    M7~\cite{richter2025investigating} & G & 30 & 3.50 & \textbf{0.87} & 14.10 & 4.08 & \textbf{4.48} \\
    CDiffuSE~\cite{lu2022conditional} & G & 200 & 2.46 & 0.79 & 12.60 & 3.32 & 3.53 \\
    SGMSE+~\cite{richter2023speech} & G & 30 & 2.93 & \textbf{0.87} & \textbf{17.30} & \textbf{4.12} & 4.24 \\
    CRP$^{\dag}$~\cite{CRP2024Lay} & G & 1 & 2.31 & 0.85 & 11.50 & 3.62 & 3.94 \\      
    \midrule  
    Teacher$^{*}$ & G & 30 & 2.90 & 0.85 & 16.90 & 3.99 & \textbf{4.19} \\
    Teacher$^{*}$ & G & 10 & 2.60 & 0.84 & 17.00 & 3.75 & 4.06 \\
    Teacher$^{*}$ & G & 5 & 1.65 & 0.74 & 12.90 & 2.47 & 3.0 \\
    Ours(Euler) & G & 1 & 2.46 & 0.83 & 14.30 & 3.27 & 3.46 \\
    Ours(Euler + RCD) & G & 1 & 2.88 & \textbf{0.87} & 18.30 & \textbf{4.05} & 4.18 \\    
    Ours(Heun) & G & 1 & 2.71 & 0.86 & 18.20 & 4.02 & 4.11 \\
    Ours(Heun + RCD) & G & 1 & \textbf{2.98} & 0.86 & \textbf{18.40} & 4.01 & 4.13 \\
    \midrule    
    Use Heun + RCD&   &   &   &   &   &   &   \\
    Ours(+ PESQ loss) & G & 1 & \textbf{3.99} & 0.83 & 0.40 & 2.93 & 2.31 \\
    Ours(+ SI-SDR loss) & G & 1 & 3.00 & \textbf{0.87} & 17.30 & 4.05 & 3.57 \\
    ROSE-CD & G & 1 & 3.49 & \textbf{0.87} & \textbf{17.80} & \textbf{4.13} & \textbf{4.41} \\
    \bottomrule
    \multicolumn{8}{l}{\footnotesize{$^{\dag}$: test with checkpoint from~\cite{CRP2024Lay}.}} \\    
    \multicolumn{8}{l}{\footnotesize{$^{*}$: teacher model are trained with the same backbone as used in SGMSE+~\cite{richter2023speech}.}} \\
\end{tabular}

}
\end{table}

\subsection{Robustness evaluation}
\label{subsec:Robustness}
We began by evaluating the generalization capability of our approach on the out-of-domain TIMIT+NOISE92 dataset. As shown in~\Cref{tab:eval_timit}, our model consistently outperforms the teacher in PESQ, SI-SDR, and WV-MOS, indicating improvements in both perceived speech quality and waveform fidelity, while maintaining comparable performance on ESTOI and MOS-SSL, thereby demonstrating strong robustness to unseen noise conditions. Notably, applying RCD with only the PESQ loss yielded the highest PESQ of 3.39 but significantly reduced SI-SDR to 0.70~dB. In contrast, using only the SI-SDR loss achieved the best SI-SDR of 15.30~dB with competitive PESQ. ROSE-CD effectively balanced the trade-off and delivered robust overall performance, achieving the highest WV-MOS score of 3.77.

\begin{table}[htbp]
\centering
\caption{Out-of-domain test results on TIMIT+NOISE92 with model trained on VB-DMD.}

\label{tab:eval_timit}
\sisetup{
    reset-text-series = false, 
    text-series-to-math = true, 
    mode=text,
    tight-spacing=true,
    round-mode=places,
    round-precision=2,
    table-format=2.2,
    table-number-alignment=center
}
\resizebox{\linewidth}{!}{
\begin{tabular}{l c c S[table-format=1.2] S[table-format=1.2] S[table-format=2.2] S[table-format=1.2] S[table-format=1.1]}
    \toprule
    \textbf{Method} & \textbf{Type} & \textbf{Steps} & \textbf{PESQ} & \textbf{ESTOI} & \textbf{SI-SDR} & \textbf{MOS-SSL} & \textbf{WV-MOS} \\
    \midrule
    SGMSE+$^{\dag}$~\cite{richter2023speech} & G & 30 & 2.08 & 0.80 & 13.90 & \textbf{3.58} & 3.72 \\
    Teacher & G & 30 & 2.16 & \textbf{0.81} & 13.90 & 3.45 & 3.73 \\
    Ours(w/o RCD) & G & 1 & 1.96 & 0.80 & 14.80 & 3.39 & 3.20 \\
    Ours(+ RCD) & G & 1 & 2.28 & \textbf{0.81} & 14.50 & 3.35 & 3.50 \\
    Ours(RCD + PESQ loss) & G & 1 & \textbf{3.40} & 0.76 & 1.00 & 1.98 & 2.11 \\
    Ours(RCD + SI-SDR loss) & G & 1 & 2.32 & \textbf{0.81} & \textbf{15.30} & 1.87 & 3.61 \\
    ROSE-CD & G & 1 & 2.62 & \textbf{0.81} & 14.70 & 3.33 & \textbf{3.77} \\
    \bottomrule
    \multicolumn{8}{l}{\footnotesize{$^{\dag}$: test with checkpoint from~\cite{richter2023speech}.}} \\
\end{tabular}
}
\end{table}

We further assessed the real-world noise robustness of our model using the DNS Challenge 2020 dataset and reference-free metrics, as shown in~\Cref{tab:eval_dns}. ROSE-CD demonstrated strong performance under practical noisy conditions, achieving a DNSMOS of 3.53 (vs. 3.62), a WV-MOS of 2.51 (vs. 2.61), and a SIG (speech quality) of 4.01 (vs. 4.08), closely approaching the performance of teacher model and validating its perceptual quality and robustness.

\begin{table}[htbp]
\centering
\caption{Real-world recordings test results on DNS Challenge 2020 with model trained on VB-DMD. Teacher model uses 30 steps.}

\label{tab:eval_dns}
\sisetup{
    reset-text-series = false, 
    text-series-to-math = true, 
    mode=text,
    tight-spacing=true,
    round-mode=places,
    round-precision=2,
    table-format=2.2,
    table-number-alignment=center
}
\resizebox{\linewidth}{!}{
\begin{tabular}{l c c S[table-format=1.2] S[table-format=1.2] S[table-format=2.2]}
    \toprule
    \textbf{Method} & \textbf{DNSMOS} & \textbf{SIG} & \textbf{BAK} & \textbf{OVRL} & \textbf{WV-MOS} \\
    \midrule    
    Conv-TasNet~\cite{luo2019conv} & 3.07 & 2.87 & 3.59 & 2.52 & 2.07 \\
    MetricGAN+~\cite{fu21_interspeech} & 3.26 & 2.88 & 3.39 & 2.45 & 1.52 \\
    SGMSE+$^{\dag}$~\cite{richter2023speech} & \textbf{3.65} & \textbf{4.10} & \textbf{4.02} & \textbf{3.66} &  \textbf{2.53} \\
    \midrule   
    Teacher & \textbf{3.62} & \textbf{4.08} & \textbf{3.89} & \textbf{3.57} & \textbf{2.61} \\
    Ours(w/o RCD)  & 3.36 & 3.67 & 3.32 & 3.03 & 2.28 \\
    Ours (+ RCD)& 3.53 & 3.88 & 3.43 & 3.19 & 2.38 \\
    Ours (RCD + PESQ loss)& 3.13 & 3.50 & 3.74 & 3.02 & 1.81 \\
    Ours (RCD + SI-SDR loss)& 3.52 & 3.95 & 3.52 & 3.27 & 2.43 \\
    ROSE-CD& 3.53 & 4.01 & 3.77 & 3.42 & 2.51 \\
    \bottomrule
    \multicolumn{6}{l}{\footnotesize{$^{\dag}$: test with checkpoint from~\cite{richter2023speech}.}} \\    
\end{tabular}
}
\end{table}

\subsection{Efficiency evaluation}
\label{subsec:Efficiency}

In~\Cref{tab:efficiency}, we report the real-time factor (RTF) for various methods. For Thunder~\cite{trachu24_interspeech}, the RTF was measured based on the StoRM~\cite{lemercier2023storm} single-step generator, since both methods share the same network architecture and adopt a two-stage framework comprising a prediction module followed by a generation module. ROSE-CD achieved a 54$\times$ speedup over the teacher model, owing to both fewer reverse steps and a simplified sampling strategy, whereas the teacher still relies on costly predictor-corrector samplers~\cite{song2021scorebased}. Moreover, compared to the hybrid approach Thunder~\cite{trachu24_interspeech}, our model operated at twice the speed, as it eliminated the need for a separate predictive network.

\begin{table}[htbp]
\centering
\caption{Performance on the VB-DMD when varying the sampling steps (RTF reported on a single NVIDIA RTX 6000).}

\label{tab:efficiency}
\sisetup{
    reset-text-series = false,
    text-series-to-math = true,
    mode=text,
    tight-spacing=true,
    round-mode=places,
    round-precision=2,
    table-format=2.2,
    table-number-alignment=center
}
\resizebox{\linewidth}{!}{
\begin{tabular}{l c c c S[table-format=1.2] S[table-format=1.2] S[table-format=2.2]}
    \toprule
    \textbf{Method} & \textbf{RTF} & \textbf{Type} & \textbf{Steps} & \textbf{PESQ} & \textbf{ESTOI} & \textbf{SI-SDR} \\
    \midrule
    StoRM~\cite{lemercier2023storm} & 3.43 & P+G & 51 & 2.93 & \textbf{0.88} & 18.80 \\
    Thunder~\cite{trachu24_interspeech} & 0.108 & P+G & 2 & \textbf{3.02} & 0.87 & \textbf{19.40} \\
    \midrule
    Teacher & 2.60 & G & 30 & 2.90 & 0.85 & 16.90 \\
    ROSE-CD & 0.048 & G & 1 & \textbf{3.49} & \textbf{0.87} & \textbf{17.80} \\
    \bottomrule
\end{tabular}
}
\end{table}

\section{CONCLUSION}
\label{sec:CONCLUSION}

This paper presents ROSE-CD, a novel one-step speech enhancement framework that leverages robust consistency distillation to achieve state-of-the-art performance. By integrating randomized learning trajectories and joint optimization of time-domain PESQ and SI-SDR losses, ROSE-CD enhances robustness, mitigates teacher model biases, and delivers superior speech quality. Evaluations on the VoiceBank-DEMAND dataset demonstrate that ROSE-CD surpasses its 30-step teacher model, achieving a PESQ score of 3.99 and a 54$\times$ inference speedup. Robustness is further validated through strong generalization on the out-of-domain TIMIT+NOISE92 dataset and real-world DNS Challenge 2020 recordings, underscoring ROSE-CD's potential for efficient, high-quality speech enhancement in practical applications.




\clearpage
\bibliographystyle{IEEEtran}
\bibliography{refs25}







\end{document}